\documentclass[10pt]{article}

\usepackage{amssymb,latexsym,amsmath}

\begin{document}

\sloppy
\renewcommand{\theequation}{\arabic{section}.\arabic{equation}}
\thinmuskip = 0.5\thinmuskip    
\medmuskip = 0.5\medmuskip
\thickmuskip = 0.5\thickmuskip
\arraycolsep = 0.3\arraycolsep

\newtheorem{theorem}{Theorem}[section]
\newtheorem{corollary}[theorem]{Corollary}
\newtheorem{lemma}[theorem]{Lemma}
\renewcommand{\thetheorem}{\arabic{section}.\arabic{theorem}}

\newcommand{\prf}{\noindent{\bf Proof.}\ }
\def\prfe{\hspace*{\fill} $\Box$

\smallskip \noindent}

\def\be{\begin{equation}}
\def\ee{\end{equation}}
\def\bea{\begin{eqnarray}}
\def\eea{\end{eqnarray}}
\def\beas{\begin{eqnarray*}}
\def\eeas{\end{eqnarray*}}

\newcommand{\R}{\mathbb R} 
\newcommand{\N}{\mathbb N}

\def\supp{\mathrm{supp}\,} 
\def\vol{\mathrm{vol}\,} 
\def\sign{\mathrm{sign}\,}
\def\ekin{E_\mathrm{kin}}
\def\epot{E_\mathrm{pot}}

\def\C{{\cal C}}
\def\H{{\cal H}}
\def\Hc{{{\cal H}_C}}

\title{Existence and stability of static shells for the Vlasov-Poisson system with a fixed central point mass}
\author{ Achim Schulze\\
         Mathematisches Institut der
         Universit\"at Bayreuth\\
         D 95440 Bayreuth, Germany}
\maketitle

\begin{abstract}
We consider the Vlasov-Poisson system with spherical symmetry and an exterior potential which is induced by a point mass in the center. This system can be used as a simple model for a newtonian galaxy surrounding a black hole.  For this system, we establish a global existence result for classical solutions with shell-like initial data, i.e. the support of the density is bounded away from the point mass singularity. We also prove existence and stability of stationary solutions which describe static shells, where we use a variational approach which was established by Y.~Guo and G.~Rein.
\end{abstract}

\section{Introduction}
\setcounter{equation}{0}
In stellar dynamics, the evolution of a large ensemble of particles (e.g. stars) which interact only by their self-consistent, self-generated gravitational field, is described by the Vlasov-Poisson system
\be
\partial_{t}f+v\cdot \nabla_{x}f-\nabla_{x}U\cdot \nabla_{v}f =0,
\label{vlasov1}
\ee
\be
\Delta U = 4\pi \rho,\ \lim_{|x| \to \infty} U(t,x) = 0, \label{poisson}
\ee
\be
\rho(t,x) = \int f(t,x,v)dv \label{rhodef}.  
\ee
Here $f= f(t,x,v) \geq 0$ is the phase-space density of the particles, where $t\in\R$ denotes time, and $x,v\in \R^3$ denote position and velocity. 
$U=U(t,x)$ is the gravitational potential of the ensemble, and $\rho=\rho(t,x)$
is its spatial density.

We want to investigate this system under the influence of a fixed point mass. If we assume that a point mass $M_c$ is fixed in the origin and acts like an external force on the particles, the Vlasov equation reads
\be
\partial_{t}f+v\cdot \nabla_{x}f-(\nabla_{x}U - \nabla_{x}\frac{M_c}{|x|})\cdot \nabla_{v}f =0.
\label{vlasov}
\ee
If we write $U_{\text{eff}} := U - M_c/|x|$,
the Poisson equation becomes
\be \label{poisson2}
\Delta U_{\text{eff}} = 4\pi \left( \rho + M_c\delta \right), 
\ee
where $\delta$ denotes the $\delta$-distribution. Here, we want examine to the existence and stability of steady states of the system (\ref{poisson})--(\ref{vlasov}). 
One easily verifies that $f=f(t,x,v)$ is a solution of (\ref{vlasov}), iff it is constant along solutions
of the characteristic system
\begin{equation} \label{charsys}
\left\{ \begin{array}{lll} \dot{X} &= &V \\ \dot{V} &=  &- \nabla _x \big( U(s,X) - \frac{M_c}{|X|} \big) \end{array} \right. ,
\end{equation}
where $(X,V)=(X,V)(s):=(X,V)(s,t,x,v)$ with $(X,V)(t,t,x,v)=(x,v)$ for an initial value $(x,v) \in \mathbb{R}^6$ and $s,t \in \mathbb{R}$.
Thus for the construction of stationary solutions, a natural idea is to find conserved quantities of (\ref{charsys}) -- now with time-independent $U$.
One immediate expression for such a quantity is the particle energy
$$ E = \frac{1}{2}|v|^2 + U(x) - \frac{M_{c}}{|x|}$$ and if we make some additional symmetry assumptions on the potential $U$, we can find other conserved terms such as the angular momentum.

In this paper, we are interested in stationary solutions of the form
\be
f_0(x,v) = (E_0 - E)_{+}^k (L - L_0)_+^l,
\label{ansatz}
\ee
where $0<k\leq l$, $(\cdot )_+$ denotes the positive part and $E_0<0, L_0>0$ are constants. $E$ is the particle energy as above and
\be
L=|x \times v|^2 = |x|^2|v|^2 - (x\cdot v)^2
\label{defl}
\ee
denotes the modulus of angular momentum squared which is conserved along characteristics, if $U$ is spherically symmetric.

If we want to construct the stationary solution $(f_0,U_0)$ explicitely from the ansatz (\ref{ansatz}), we still have to solve the Poisson equation (\ref{poisson}) to get a self-consistent potential $U_0$. The existence of stationary solutions with parameter range $k>-1, \enspace l>-1, \enspace k+l+1/2\geq0, \enspace k<3l+7/2$ was established in \cite{REINEX} for $M_c = 0$.

Without the exterior potential, the existence and stability of stationary solutions of the form (\ref{ansatz}) was done in \cite{SCH}, where the parameter range $l>-1$ and $0<k<l+3/2$ was covered. 

As mentioned above, for our ansatz (\ref{ansatz}) we require that the corresponding potential $U$ is spherically symmetric and therefore the  stationary solutions (\ref{ansatz}) also have to be spherically symmetric, i.e.,
\be
f(x,v) = f(Ax,Av) \quad \forall A \in O(3),
\label{sphsymm}
\ee
where $O(3)$ is the group of orthogonal $3\times 3$ matrices. For $L_0 >0$ the support of the induced spatial density $\rho (x) = \rho(|x|)$ is contained in some interval $[R_1,R_2]$ with $R_1 > 0$ and the steady state describes a shell. This can be seen as follows. If we introduce the new coordinates $r:= |x|, \, w:= x\cdot v/r$ and $L$ as in (\ref{defl}), we can calculate the spatial density of $f_0$ as
\begin{align}
\rho_{f_0}(x) &= \int_{\mathbb{R}^3} f_0(x,v) \, dv \notag \\
&= \frac{\pi}{r^2} \int_{\mathbb{R}} \int_0^{\infty} \bigg( E_0 -\frac{1}{2}\big( w^2 + \frac{L}{r^2}\big) - U_0(r) + \frac{M_c}{r} \bigg)_+^k \, (L-L_0)_+^l \, dwdL \label{rwl} \\
&= C(k,l) \, r^{2l} \, \bigg( E_0 - U_0(r) + \frac{M_c}{r} - \frac{L_0}{r^2} \bigg)_+^{k+l+3/2} \label{rwl2}, 
\end{align}
where 
$$C(k,l) =  2^{l+3/2} \pi \int_0^1 \frac{s^l}{\sqrt{1-s}} \, ds \,  \int_0^1 s^{l+1/2} (1-s)^k \, ds. $$
For small $r$ the expression in the bracket of (\ref{rwl2}) becomes negative and this implies $\supp \, \rho_f \subset [R_1,\infty[$, for some $R_1 >0$. On the other hand, because of $U_0'(r) = \int_0^r s^2 \, \rho(s) \, ds /r^2 >0$ for $ r >0$, the function $-U_0(r) + M_c/r$ is decreasing and with $E_0 <0$ we conclude that $\rho_{f_0} (r) = 0$ for large $r$. 

These shells together with the exterior potential induced by a point mass can be used as a simple model for a galaxy which encloses a black hole in the center. 

The ansatz (\ref{ansatz}) also leads to steady states and shells of the Vlasov-Einstein system, the general relativistic counterpart of the newtonian Vlasov-Poisson system, and they provide an access to study stability and critical phenomena numerically, cf. \cite{REINAND}. 

We examine the shells in the newtonian framework and to investigate their stability, we will firstly prove a global existence result for the system (\ref{poisson})--(\ref{vlasov}) for initial data, which vanishes in a neighbourhood of the singularity $r=0$. The corresponding solution then exists for all time, and will always vanish, if $x$ is in a ball around the singularity, which is determined by the initial datum.  We mention that, without the exterior potential, the existence problem for the Vlasov--Poisson system is well understood, see for example \cite{LP,Pfaffel,Schaeff} for global existence of classical solutions. However, in our situation the exterior potential becomes unbounded in $r=0$ and we have to ensure that the particles stay away from the singularity.

To show existence and stability of the shells, we use a similar approach as in \cite{REINST}, where existence and stability of the above steady states was shown in the case $L_0 = 0$ without the exterior potential. The main idea is to use an Energy-Casimir functional as a Lyapunov function with the help of variational methods. Concerning this approach for stability issues for the Vlasov--Poisson system we also want to list \cite{GUO1,GUO2,REIN15,REIN16,REIN12,REIN11,REIN13} here. We briefly sketch the basic concept:

The Vlasov-Poisson system is conservative, i.e., the total energy
\begin{align}
\H (f) &:= \ekin (f) + \epot (f) \notag\\
&:= \frac{1}{2} \int |v|^2 f(x,v)\,dv\,dx - \frac{1}{8 \pi} \int \left( |\nabla U_f(x)|^2 + \frac{8\pi M_c}{|x|} \rho_f (x) \right) dx
\label{defenergy}
\end{align}
of a state $f$ is conserved along solutions and hence is a natural candidate for a 
Lyapunov function in a stability analysis; $U_f$ denotes the potential
induced by $f$, note also the interaction term $\int \rho_f \, M_c/|x| \, dx$ induced by the fixed central point mass. However, the energy does not have critical
points, but
for any reasonable function $\Phi$ the so-called {\em Casimir functional}
\[
\C(f) := \iint \Phi(f(x,v))\,dv\,dx
\]
is conserved as well. Now one tries to minimize the energy-Casimir functional 
\[
\Hc := \H + \C
\]
in the class of allowed perturbations $\mathcal{F}_M$, which consists of positive $L^1(\mathbb{R}^6)$-functions with prescribed mass $M$, i.e. $\iint f =M$ and with finite kinetic energy and a finite Casimir functional to ensure that $\Hc$ is well-defined. 

The aim is to prove that a minimizer $f_0$ is a stationary solution of (\ref{poisson})--(\ref{vlasov}) and to deduce its stability. One of the difficulties is to show that the weak limit of a minimizing sequence in $\Hc$, indeed is a minimizer. For this purpose, we will need that every function in the class of perturbations $\mathcal{F}_M$ vanishes on the set $0 \leq  L <L_0$.

We are only able to show stability against spherically symmetric perturbations, because our approach requires an $L$-dependence in the Casimir functional, more precisely, we define
\be
\mathcal{C}(f) :=  \int_{\mathbb{R}^3} \Phi\left( (L-L_0)_+^{-l} f(x,v) \right) (L-L_0)_+^l \, dvdx,
\label{defcas1}
\ee
with $0<k\leq l$ as in (\ref{ansatz}), $\Phi$ convex, satisfying certain growth conditions, and this will be a conserved quantity for spherically symmetric $f$ only.
To simplify our presentation, we focus on the case 
\[
\Phi(f) = f^{1+1/k}
\]
which will lead to stationary solutions of the form (\ref{ansatz}). The Casimir functional then reads
\be
\mathcal{C}(f) :=  \int_{\mathbb{R}^3} f^{1+1/k}(x,v) (L-L_0)_+^{-l/k} \, dvdx.
\label{defcas2}
\ee
At one point we need a scaling argument, which gets complicated in the case of a translation in $L$ in the Casimir-functional. Here we exploit the spherical symmetry and use coordinates adapted to it: If $f=f(x,v)$ is spherically symmetric, we have
\[
f(x,v) = \tilde{f} (r,w,L),
\] 
with $r=|x|, \, w= \frac{x\cdot v}{r}$ and $L$ as in (\ref{rwl}), see Section 1.4.
Altogether, we want to minimize the energy-Casimir functional
$$ \Hc (f)  = \ekin (f) + \epot (f) + \mathcal{C} (f), $$
with $\ekin, \, \epot$ from (\ref{defenergy}) and $\mathcal{C} (f)$ as in (\ref{defcas2}) over the set
\begin{align} \label{defoffm}
\mathcal{F}_M := \bigg\{  &f \in L^1(\mathbb{R}^6 ) \enspace | \enspace f\geq0, \enspace f \enspace \text{is spherically symmetric,} \enspace \iint f = M, \notag \\ &\quad \ekin (f) + \mathcal{C} (f) < \infty, \enspace 
 f(x,v) = 0 \enspace \text{a.e.} \enspace \text{for} \enspace 0 \leq L < L_0 \bigg\}.
\end{align}
See (\ref{sphsymm}) for the definition of spherical symmetry.

This paper is organized as follows: In the next section, we prove a global existence result for the system (\ref{poisson})--(\ref{vlasov}). Afterwards, we examine the variational problem and we show that $\Hc$ is bounded from below in Section 1.3. Then we prove a scaling property and that the the infimum of $\Hc$ is negative in Section 1.4. In Sections 1.5 and 1.6 we show the existence of a minimizer and analyse its properties; it is a stationary solution, and it is nonlinearly stable against sperically symmetric perturbations. 

\section{Global existence}
\setcounter{equation}{0}
In order to prepare the stability analysis, we want to prove a global existence result for classical solutions to the system (\ref{poisson})--(\ref{vlasov}), so that we know that solutions in a neighbourhood of the examined steady states exist.
We want to prove the following theorem.
\begin{theorem}
Consider the system (\ref{poisson})--(\ref{vlasov}). Let $\mathring{f} \in C_c^1$ be a spherically symmetric initial datum with $\mathring{f} (x,v) =0$ for $L:= |x \times v |^2 \leq L_0$, where $L_0>0$ is given. Then the corresponding solution $(f,U)$ exists for all time and there exists $R_{\text{min}}>0$, such that $f(t,x,v) = 0$ for $|x| < R_{\text{min}}, \enspace t\geq0$, where the number $R_{\text{min}}$ only depends on $M_c, L_0$ and $\mathring{f}$.
\end{theorem}
\textbf{Remark.} Without the exterior potential, the global existence result was proved by J. Batt, cf. \cite{Batt1} and this was also the first global existence result for the Vlasov--Poisson system in space dimension three. In our proof given below, the main idea for proving the boundedness of the velocities, is due to E. Horst, cf. \cite{Horst1}.
\vspace{5mm}

\prf
We fix an initial datum $\mathring{f} \in C_c^1(\mathbb{R}^6)$ with $\mathring{f} \geq 0$ and we fix $\mathring{R}, \mathring{P}$ with 
$$ \mathring{f} (x,v) = 0 \quad \text{for} \, \, |x| \geq \mathring{R} \enspace \text{or} \enspace |v| \geq \mathring{P}. $$
This implies $\mathring{f} (x,v)= 0$ for $|x| < \sqrt{L_0}/\mathring{P}$, since $L= \sin^2(\alpha) |v|^2|x|^2 >L_0$ on the support of $\mathring{f}$, where $\alpha$ denotes the angle between $x$ and $v$. \\
In the following, we will denote first partial derivatives with respect to $x$ with $\nabla _x$ and we will write $\partial _x ^2$ for the second partial derivatives.
We now consider the following iteration process to construct the classical solution. The 0th iterate is defined by
$$ f_0(t,z) := \mathring{f} (z), \quad t\geq0, \, \, z \in \mathbb{R}^6. $$
If the $n$th iterate $f_n$ is already defined, we define
$$ \rho _n:= \rho _{f_n}:= \int_{\mathbb{R}^3} f_n \, dv, \quad U_n:= U_{\rho _n}:= -\rho_n * \frac{1}{|\cdot |}, \quad U_{n,\text{eff}}:= U_{\rho _n} - \frac{M_c}{ |\cdot|}$$
on $[0,\infty[ \times \mathbb{R}^3$, and for $L = |x\times v|^2 >L_0$ we denote by
\be \label{char1}
 Z_n (s,t,z) := (X_n,V_n)(s,t,x,v)
\ee
the solution of the characteristic system
\be \label{char2}
 \dot{X} =V, \quad \dot{V} = - \nabla _x U_{n,\text{eff}} (s,X)
\ee
with $Z_n (t,t,z) = z$, where we want to examine characteristics which start on the support of $\mathring{f}$. We claim that $|X_n(s,0,z)|$ is bounded from below by a positive constant for all $s\geq0, \, n\in \mathbb{N}$, so that the right-hand side of the charcteristic system is well-defined for all time. Together with (\ref{char1})--(\ref{char2}) this leads to the definition
\[
 f_{n+1} (t,x,v) := \left\{ \begin{array}{lll} \mathring{f} (Z_n(0,t,z)) & \text{for} \enspace z=(x,v):\enspace  |x\times v|^2 >L_0 \\ 0  & \text{else} . \end{array} \right.
\]
for the $(n+1)$st iterate. Note that, due to sperical symmetry, $L=|X\times V|^2$ is a conserved quantity of (\ref{char2}) and that $\|f_n(t)\|_1= \|\rho_n(t)\|_1 = \|\mathring{f}\|_1$ since the characteristic flow is measure preserving.
We introduce some notations:
\begin{align*}
P_0(t)&:= \mathring{P}, \\
P_n(t) &:= \sup \left\{ \left| V_{k-1} (s,0,z) \right| \enspace | \enspace  z \in \supp \mathring{f}, \enspace 0\leq s \leq t, \enspace 1\leq k\leq n \right\}, \quad n \in \mathbb{N}, \\ 
R_{\text{min}}^0(t)&:=  \sqrt{L_0}/\mathring{P}, \\
R_{\text{min}}^n(t) &:= \inf \left\{ |x|, \enspace (x,v) \in \mathbb{R}^3\times \mathbb{R}^3 \enspace | \enspace (x,v )\in \supp \, f_n(s),  \enspace 0\leq s \leq t \right\} \\
&= \inf \left\{ \left|  X_{n-1} (s,0,z) \right| \enspace | \enspace z \in \supp \mathring{f}, \enspace 0\leq s \leq t \right\}, \quad n \in \mathbb{N}. 
\end{align*}
Next we show 
$$P_n(t) \leq C_0,\enspace R_{\text{min}}^n(t)\geq C_1,  \quad t \in \mathbb{R}^+, \quad n\in \mathbb{N},$$ 
where $C_0>0$ and $C_1>0$ only depend on $M_c,\enspace \| \mathring{f}\|_1, \enspace \| \mathring{f} \|_{\infty}$ and $L_0$. 
We abbreviate $X_n(s) := X_n(s,0,z)$ for some $z \in \supp \mathring{f}$ fixed. Now fix $t>0$; we then have $X_n \in C^2([0,t])$ with
\be \label{ddotxn}
\ddot{X}_n (s) = -\left(m_{\rho_n}(s,|X_n(s)|)+M_c\right) \cdot \frac{X_n(s)}{|X_n(s)|^3},
\ee
where we used the spherical symmetry and defined $$ m_{\rho_n} (s,r) := 4\pi \int_0^r \tau^2\rho_{n}(s,\tau)\, d\tau .$$
To get suitable bounds for the right-hand side of equation (\ref{ddotxn}), we firstly use \cite{BIBEL}, Lemma P1:
\begin{align*}
\frac{m_{\rho_n}(s,|X_n(s)|)}{|X_n(s)|^2} &= |\nabla _x U_n (s,X(s)) | \\
&\leq 3(2\pi)^{2/3} \| \rho_n(t)\|_1^{1/3} \|\rho_n(t)\|_{\infty}^{2/3} \\
&\leq 4\cdot 3^{1/3} \pi^{4/3} \|\mathring{f} \|_1^{1/3} \|\mathring{f} \|_{\infty}^{2/3} P^2_n(t) \\
&=: \kappa P^2_n(t).
\end{align*}
Furthermore, since $L_n(s):= \left| X_n(s) \times V_n(s) \right|^2$ is constant in $s$,
$$ |X_n(s)|^2 \geq \frac{L_0}{|V_n(s)|^2|\sin^2(\angle(X_n(s),V_n(s)))|}\geq \frac{L_0}{P^2_{n+1}(s)}, $$
which implies
$$ \bigg| \frac{X_n(s)}{|X_n(s)|^3} \bigg| \leq  \frac{1}{L_0} P^2_{n+1}(t), \quad 0\leq s \leq t. $$
We also have
$$ |\ddot{X}_n (s)| \leq \frac{ \|\mathring{f} \|_1 + M_c}{|X_n(s)|^2}. $$
Altogether,
$$ |\ddot{X}_n (s)| \leq C^* \min \bigg\{ \frac{1}{|X_n(s)|^2}, P^2_{n+1}(t) \bigg\},\quad 0\leq s \leq t, $$
where 
$$ C^* = C^* (\|\mathring{f}\|_1,\|\mathring{f}\|_{\infty},M_c,L_0) = \max \bigg( \|\mathring{f}\|_1+M_c, \frac{M_c}{L_0}+\kappa \bigg).$$
Now define $\xi _n (s):= (X_n(s))_i$ for $i\in \{1,2,3\}$ and $0\leq s \leq t$. Then
$$ |\ddot \xi_n (s)| \leq g(\xi (s)), \quad 0\leq s \leq t, $$ 
where 
$$ g(r) := C^*\min \bigg\{ \frac{1}{r^2}, P^2_{n+1}(t) \bigg\}, \quad r\in \mathbb{R}.$$
If $\dot{\xi}_n (s) \neq 0$ on $]0,t[$, we have
\begin{align*}
\left| \dot{\xi}_n(t) - \dot{\xi}_n (0) \right|^2 &\leq \left| \dot{\xi}_n(t) - \dot{\xi}_n (0) \right| \,\, \left| \dot{\xi}_n(t) + \dot{\xi}_n (0) \right| \\
&= \left| \dot{\xi}_n(t)^2 - \dot{\xi}_n (0)^2\right| = 2\left| \int_0^t \dot{\xi}_n (s) \ddot{\xi}_n(s) \, ds\right| \\
&\leq 2 \int_0^t\big| \dot{\xi}_n (s)\big| g\big( \xi_n(s)\big) \, ds = 2 \int_{\xi ([0,t])} g(r)\, dr \\
&\leq  2\int_{\mathbb{R}} g(r) \, dr = 8C^* P_{n+1}(t),
\end{align*}
and hence
$$\left| \dot{\xi}_n(t) - \dot{\xi}_n (0)\right| \leq 2\sqrt{2C^*}P^{1/2}_{n+1} (t).$$
If $\dot{\xi}_n (s) = 0$ for some $s\in ]0,t[$, we define
$$ s_{-}:= \inf\{ s \in ]0,t[ \enspace | \enspace \dot{\xi}_n (s) = 0 \}, \qquad  s_{+}:= \sup\{ s \in ]0,t[ \enspace | \enspace \dot{\xi}_n (s) = 0 \} $$
and the calculation made above implies
\begin{align*}
\left| \dot{\xi}_n(t) - \dot{\xi}_n (0) \right| &\leq \left| \dot{\xi}_n(t) - \dot{\xi}_n (s_+) \right| + \left| \dot{\xi}_n(s_-) - \dot{\xi}_n (0)\right| \\
& \leq 4 \sqrt{2C^*}P^{1/2}_{n+1} (t).
\end{align*}
Since $\dot{\xi}_n = (\dot{X}_n)_i = (V_n)_i$, we conclude 
$$ P_{n+1}(t) \leq \mathring{P} + 4\sqrt{6C^*}P^{1/2}_{n+1} (t)  \qquad t\geq 0, \quad n\in \mathbb{N}. $$ 
and therefore
$$ P_n(t) \leq C_0, \quad n \in  \mathbb{N}, $$
where $C_0$ only depends on $\| \mathring{f}\|_1,\| \mathring{f}\|_{\infty},L_0, M_c$ and we also have 
$$R_{\text{min}}^n(t)\geq  \frac{\sqrt{L_0}}{C_0} , \quad n \in \mathbb{N}.$$
Now we can continue with the iterates and prove their convergence.
We have
$$ f_n \in C^1([0,\infty[\times \mathbb{R}^6), \quad \| f_n(t) \|_{\infty} = \|\mathring{f}\|_{\infty}, \quad \|f_n(t) \|_1 = \|\mathring{f}\|_1, \quad t\geq 0, $$
$$ f_n(t,x,v) = 0 \quad \text{for}\enspace |v| \geq P_n(t) \, \, \text{or} \, \, |x| \geq \mathring{R} + \int_0^t P_n(s) \, ds,\, \, \text{or} \, \, L \leq L_0, $$
and
\begin{align*}
&\rho _n \in C^1([0,\infty[\times \mathbb{R}^3), \\
&\|\rho_n(t)\|_1 = \|\mathring{f}\|_1, \quad \|\rho_n(t)\|_{\infty} \leq \frac{4\pi}{3} \| \mathring{f} \|_{\infty} P_n^3(t), \quad t\geq0, \\
&\rho_n (t,x) = 0\quad \text{for} \, \, |x| \geq \mathring{R} + \int_0^t P_n(s) \, ds 
\end{align*}

We define
$$ \| \nabla_x U_{n,\text{eff}}(t) \|_{\text{min},\infty} := \sup \left\{ |\nabla_x U_{n,\text{eff}}(t,x)| \, \, | \, \, \frac{\sqrt{L_0}}{C_0} \leq |x| < \infty \right\}$$
and $ \| \partial^2_x U_{n,\text{eff}}(t) \|_{\text{min},\infty}$ is defined analoguously. \\ \\
Now choose $T_0>0$.  We want to prove that there exists a constant $C>0$, which only depends on $T_0, \mathring{f}, L_0$ and $M_c$, such that
$$ \| \nabla_x \rho_n (t) \|_{\infty} +  \| \partial^2_x U_{n,\text{eff}}(t) \|_{\text{min},\infty} \leq C, \quad t\in [0,T_0], \, \, \, n\in \mathbb{N} .$$
In the following, $C>0$ may change from line to line, but there is no dependence on $t\in [0,T_0]$ or $n\in \mathbb{N}$. We have
$$ \left| \nabla _x \rho_{n+1} (t,x)\right| \leq \int_{|v|\leq P_n(t)} \left| \nabla_x \left[ \mathring{f} \left(Z_n(0,t,x,v)\right) \right] \right| \, dv \leq C \| \nabla _x Z_n(0,t,\cdot)\|_{\infty}^*, $$
where 
$$
 \| \nabla _x Z_n(0,t,\cdot)\|_{\infty}^* := \sup \left\{ \big| \nabla _x Z_n(0,t,z) \big| \enspace | \enspace z: \, Z(0,t,z) \in \supp \mathring{f} \right\}. $$
Next, fix $x,v \in \mathbb{R}^3$, $t\in [0,T_0]$ and write $Z_n(s)= \left( X_n,V_n \right) (s) := \left( X_n,V_n \right) (s,t,x,v)$, where we require that $Z_n(0) \in \supp \mathring{f}$.
Differentiating the characteristic system with respect to $x$, we get
$$ |\nabla_x \dot{X}_n(s)| \leq |\nabla _x V_n(s)|, \quad |\nabla_x \dot{V}_n(s)|\leq \| \partial_x^2 U_{n,\text{eff}}(s)\|_{\text{min},\infty} |\nabla _x X_n(s)|. $$
By integrating and noticing $\nabla _x X_n(t) = E, \nabla _x V_n(t) = 0$, we have
\begin{align*}
\big| \nabla_x &X_n(s) \big| +\big| \nabla_x V_n(s) \big| \\
& \leq 1+\int_s^t \big( 1+  \| \partial^2_x U_{n,\text{eff}}(\tau) \|_{\text{min},\infty} \big)\, \, \big( \big| \nabla _x X_n(\tau) \big| +\big| \nabla _x V_n(\tau) \big| \big) \, d\tau.
\end{align*}
Gronwall's lemma now implies
$$\big| \nabla_x X_n(s) \big| +\big| \nabla_x V_n(s) \big| \leq \exp \int_0^t  \big( 1+  \| \partial^2_x U_{n,\text{eff}}(\tau) \|_{\text{min},\infty} \big)\, \, d\tau, $$
and thus
$$ \| \nabla_x \rho_{n+1} (t) \|_{\infty} \leq C\exp \int_0^t  \| \partial^2_x U_{n,\text{eff}}(\tau) \|_{\text{min},\infty}\, \,d\tau. $$
A well known estimate for the Poisson equation then implies, cf.\cite{BIBEL}, Lemma P1,
$$  \| \partial^2_x U_{n,\text{eff}}(t) \|_{\text{min},\infty}  \leq C  \big( 1+  \int_0^t\| \partial^2_x U_{n,\text{eff}}(\tau) \|_{\text{min},\infty} \,\, d\tau \big). $$
By induction, 
$$\| \partial^2_x U_{n,\text{eff}}(t) \|_{\text{min},\infty} \leq Ce^{Ct}$$ 
and thus $\| \partial^2_x U_{n,\text{eff}}(t) \|_{\text{min},\infty}  \leq C $.
Now we show that the sequence $(f_n)$ converges to some function $f$, uniformly on $[0,T_0]\times \mathbb{R}^3 \times \mathbb{R}^3$. For $n \in \mathbb{N}$ and $z \in \mathbb{R}^6$,
$$ |f_{n+1}(t,z) - f_n(t,z)| \leq C|Z_n(0,t,z) - Z_{n-1}(0,t,z)| .$$ 
For $0\leq s \leq t$, we have
\begin{align*}
|X_n(s) - X_{n-1}(s)| &\leq \int_s^t |V_n(\tau)-V_{n-1}(\tau)|d\tau , \\
|V_n(s) - V_{n-1}(s)| &\leq \int_s^t \bigg[ |\nabla_x U_n(\tau,X_n(\tau))- \nabla_x U_{n-1}(\tau,X_n(\tau))| \\
& \hspace{10mm} + |\nabla_x U_{{n-1},\text{eff}}(\tau,X_n(\tau))- \nabla_x U_{{n-1},\text{eff}} (\tau,X_{n-1}(\tau))| \bigg] \, d\tau \\
&\leq \int_s^t \bigg[ \big\| \nabla_x U_n(\tau)- \nabla_x U_{n-1}(\tau) \big\| _{\infty} \\
& \hspace{10mm} + 2\| \partial^2_x U_{n-1,\text{eff}} (\tau)\|_{\text{min},\infty} |X_n(\tau) - X_{n-1}(\tau) | \bigg] \,  d\tau,
\end{align*}
where we used the mean value theorem and the factor $2 |X_n(\tau) - X_{n-1}(\tau) |$ in the last line is an upper bound for the length of a curve which connects $X_n(\tau)$ with $X_{n-1}(\tau)$ ($s\leq \tau \leq t$) and avoids the critical area $B_{\sqrt{L_0}/C_0}$ -- note again that we have the inequality $R^n_{\text{min}} (t) \geq \sqrt{L_0}/C_0$.

Recalling  $\| \partial^2_x U_{n,\text{eff}}(t) \|_{\text{min},\infty}  \leq C $, adding these estimates and applying Gronwall's lemma, we obtain
\begin{align*}
|Z_n (s) - Z_{n-1} (s) | &\leq C \int_s^t \big\| \nabla_x U_n(\tau)- \nabla_x U_{n-1}(\tau) \big\| _{\infty} \, d\tau \\
&\leq C \int_0^t  \big\| \rho_n(\tau)- \rho_{n-1}(\tau) \big\| _{\infty}^{2/3}  \big\| \rho_n(\tau)- \rho_{n-1}(\tau) \big\| _1^{1/3} \, d\tau \\
&\leq C \int_0^t  \big\| \rho_n(\tau)- \rho_{n-1}(\tau) \big\| _{\infty}\, d\tau \\
&\leq C \int_0^t  \big\| f_n(\tau)- f_{n-1}(\tau) \big\| _{\infty}, 
\end{align*}
where the second inequality follows by splitting the expression
$$ | \nabla_x U(x) | \leq \int \frac{\rho (x)}{|x-y|^2} \, dy \leq  \int_{|x-y|<R} \frac{\rho (x)}{|x-y|^2} \, dy 
+   \int_{|x-y|\geq R} \frac{\rho (x)}{|x-y|^2} \, dy, $$
and then using H\"older's inequality and an optimization in $R>0$, cf. \cite{BIBEL}, Lemma P1. \\
Also note that the support of both $\rho_n(t)$ and $f_n(t)$ is bounded, uniformly in $n$ and $t \in [0,T_0]$.
Altogether, we have
$$ \big\| f_{n+1}(t)- f_{n}(t) \big\| _{\infty} \leq C_* \int_0^t \big\| f_{n}(\tau)- f_{n-1}(\tau) \big\| _{\infty} \, d\tau, $$
and by induction,
$$ \big\| f_n(t)- f_{n-1}(t) \big\| _{\infty} \leq C \frac{C^n_* t^n}{n!} \leq C\frac{C^n}{n!}, \quad n\in \mathbb{N}_0, \enspace 0\leq t \leq T_0 .$$
This implies that the sequence $(f_n)$ is uniformly Cauchy and converges uniformly on $[0,T_0]\times \mathbb{R}^6$ to some function $f \in C([0,T_0]\times \mathbb{R}^6)$, which has the following property:
$$ f(t,x,v) = 0 \quad \text{for} \quad |v| \geq C_0 \quad \text{or} \quad |x| \geq \mathring{R} + C_0t. $$
Furthermore,
$$ \rho _n \rightarrow \rho := \rho _f , \quad U_n \rightarrow U := U_f, \quad (n\rightarrow \infty),$$
uniformly on $[0,T_0] \times \mathbb{R}^3$.
Since $T_0 >0$ was arbitrary, the proof is complete once we show that the limit function $f$ has the regularity to be a solution to the Vlasov--Poisson system. With \cite{BIBEL}, Lemma P1, we have
$$ \|\nabla _x U_n(t) - \nabla _x U_m (t)\|_{\infty} \leq C \| \rho _n (t) - \rho _m(t) \|_{\infty}^{2/3} \| \rho _n (t) - \rho _m(t) \|_1^{1/3} $$
and
\begin{align*}
\|\partial^2 _x U_n(t) - \partial^2 _x U_m (t)\|_{\infty} &\leq C \bigg[ \bigg( 1+ \ln \frac{R}{d} \bigg) \|\rho _n(t) - \rho _m(t) \|_{\infty} \\
&\hspace{10mm} + d   \|\nabla _x\rho _n(t) - \nabla _x\rho _m(t) \|_{\infty} + R^{-3} \| \rho _n (t) - \rho _m(t) \|_1 \bigg]
\end{align*}
for any $0<d\leq R$. This implies that the sequences $(\nabla _xU_n)$ and $(\partial _x^2 U_n)$ are also uniform Cauchy sequences on $[0,T_0] \times \mathbb{R}^3$. Indeed, since all $\rho_n$ have compact support, uniformly in $n$, we can estimate 
$$ \| \rho _n (t) - \rho _m(t) \|_1 \leq C \| \rho _n (t) - \rho _m(t) \|_{\infty} \leq C \| f _n (t) - f _m(t) \|_{\infty} $$
which converges to zero. For the term with the derivatives of $\rho _n$, we only know that
$$ \|\nabla _x\rho _n(t) - \nabla _x\rho _m(t) \|_{\infty} \leq C$$
with a not necessarily small constant $C$, but here we can choose $d>0$ in front of this term as small as we want. Hence we have
$$ \nabla _x U, \partial _x^2 U \in C ([0,T_0]\times \mathbb{R}^3).$$
Now we have for the characterstic flow $Z$, induced by the limiting field $-\nabla_x U$,
$$Z= \lim_{n\rightarrow \infty} Z_n \in C^1([0,T_0] \times [0,T_0] \times \mathbb{R}^6), $$
and finally,
$$f(t,z) = \lim_{n\rightarrow \infty} \mathring{f} \big( Z_n (0,t,z)  \big) = \mathring{f} \big( Z(0,t,z) \big), $$
so that $f \in C^1\big([0,T_0] \times \mathbb{R}^6 \big)$ is a classical solution.
\prfe
\section{A lower bound on $\Hc$}
\setcounter{equation}{0}
We recall that we want to minimize 
$$ \Hc (f)  = \ekin (f) + \epot (f) + \mathcal{C} (f), $$
with $\ekin, \, \epot$ from (\ref{defenergy}) and $\mathcal{C} (f)$ as in (\ref{defcas2}) over the set
\begin{align*} 
\mathcal{F}_M := \bigg\{  &f \in L^1(\mathbb{R}^6 ) \enspace | \enspace f\geq0, \enspace f \enspace \text{is spherically symmetric,} \enspace \iint f = M, \notag \\ &\quad \ekin (f) + \mathcal{C} (f) < \infty, \enspace 
 f(x,v) = 0 \enspace \text{a.e.} \enspace \text{for} \enspace 0 \leq L < L_0 \bigg\}.
\end{align*}
Firstly, we want to establish a lower bound on $\Hc$ and we will need several estimates for $\rho _f$ and $U_f$ induced by an element $f \in \mathcal{F}_M$. We will show that $\epot (f)$ makes sense, that is, 
$$ \nabla U_f \in L^2 (\mathbb{R}^3) \quad \text{and} \quad \int_{\mathbb{R}^3} \frac{M_c}{|x|} \rho_f(x) \, dx < \infty .$$
\begin{lemma}  \label{lemmarhobound}
Let $n := k + l + \frac{3}{2} $. Then there exists $C>0$, such that
\[
\int \rho_f^{1+\frac{1}{n}}(x) |x|^{-2l/n} dx \leq C ( \mathcal{C} (f) + \ekin(f) ), \quad f \in \mathcal{F}_M.
\]
\end{lemma}
\prf
For any $R>0$, we have
\begin{align*}
\rho_f (x) &= \int f(x,v) \, dv \\
&= \int_{ |v| \leq  R } f(x,v) \, dv + \int_{ |v| \geq R } f(x,v) \, dv \\
&\leq \int_{|v| \leq R } (L-L_0)_+^{\frac{l}{k+1}} f(x,v) (L-L_0)_+^{-\frac{l}{k+1}} dv + \frac{1}{R^2} \int |v|^2 f(x,v)\, dv \\
&\leq C  \left( \int_{|v| \leq R} (L - L_0 )_+^{l} \, dv \right)^{\frac{1}{k+1}}
\left( \int f^{1+\frac{1}{k}} (x,v) (L-L_0)_+^{-l/k} \, dv \right)^{\frac{k}{k+1}} \\
&\quad + \frac{1}{R^2} \int |v|^2 f(x,v) \, dv \\
&\leq C \, |x|^{\frac{2l}{k+1}} \, R^{\frac{2l+3}{k+1}} \left( \int f^{1+\frac{1}{k}} (x,v) (L-L_0)_+^{-l/k} \, dv \right)^{\frac{k}{k+1}} \\
&\quad + \frac{1}{R^2} \int |v|^2 f(x,v)\, dv.
\end{align*}
Optimization in $R$ yields
$$ R:=\left[  \bigg(2 \int |v|^2 f(x,v)\, dv \bigg) |x|^{\frac{-2l}{k+1}} \left( \int f^{1+\frac{1}{k}} (x,v) (L-L_0)_+^{-l/k} \, dv \right)^{\frac{-k}{k+1}}\right]^{\frac{k+1}{2l+2k+5}}, $$
and thus
\begin{align*}
\rho_f (x) &\leq C |x|^{\frac{2l}{k+l+5/2}}  \left( \int f^{1+\frac{1}{k}} (x,v) (L-L_0)_+^{-l/k} \, dv \right)^{\frac{k}{l+k+5/2}} \left( \ekin (f) \right) ^{\frac{l+3/2}{l+k+5/2}} \\
& \leq C |x|^{\frac{2l}{k+l+5/2}} \left(\ekin (f) + \int f^{1+\frac{1}{k}} (x,v) (L-L_0)_+^{-l/k} \, dv \right)^{\frac{n}{n+1}}.
\end{align*}
Taking both sides of the inequality to the power $1+\frac{1}{n}$, dividing by $r^{\frac{2l}{n}}$ and integrating 
with respect to $x$ proves the assertion.
\prfe
From Lemma \ref{lemmarhobound} we see that a function $f$ lying in $\mathcal{F}_M$ and its induced density $\rho_f$ automatically
are elements of certain Banach spaces which we now define:
\begin{align*}
L^{k,l} (\mathbb{R}^6):= \bigg\{ &f: \mathbb{R}^6 \rightarrow \mathbb{R} \enspace \text{measurable, spherically symmetric and} \\ &\iint |f|^{1+\frac{1}{k}} (L-L_0)_+^{-l/k} \, dxdv< \infty \bigg\}
\end{align*}
equipped with the norm 
$$ \| f \|_{k,l} := \left( \iint |f|^{1+\frac{1}{k}} (L-L_0)_+^{-l/k}\, dxdv\right)^{\frac{k}{k+1}} $$
and
\begin{align*}
L^{n,l} (\mathbb{R}^3):= \bigg\{ &\rho : \mathbb{R}^3 \rightarrow \mathbb{R}\enspace \text{measurable, spherically symmetric and} \\\ & \int |\rho|^{1+\frac{1}{n}} |x|^{-2l/n} dx < \infty \bigg\}
\end{align*}
with norm
$$ \| \rho \|_{n,l} := \left( \int |\rho|^{1+\frac{1}{n}} |x|^{-2l/n} dx \right)^{\frac{n}{n+1}}. $$
Both spaces are reflexive Banach spaces. More precisely, $f$ and $\rho _f$ are contained in the subsets $L^{k,l}_+(\mathbb{R}^6)$ and $L^{n,l}_+(\mathbb{R}^3)$, respectively, which consist of the a.e.-nonnegative functions of these spaces.

We now need some notations which clarify what $\epot (f)$ and $\nabla U_f$ means for $f \in \mathcal{F}_M$. 
For spherically symmetric $\rho \in C^1_c(\mathbb{R}^3)$ Poisson's equation becomes
$$ \frac{1}{r^2}\big( r^2U'(r) \big)' = 4\pi \rho(r), $$
where $r:=|x|$ and we have $U\in C^2(\mathbb{R}^3)$ and $U'(r) = 4\pi \int_0^r s^2\rho(s) \, ds /r^2$; in particular, $\nabla U(x) = U'(r) \frac{x}{r}.$
This motivates the following definitions. For $f \in \mathcal{F}_M$, i.e., $\rho:= \rho_f \in L^{n,l}(\mathbb{R}^3)$, we define
\begin{align}
m_{\rho} (r) &:= \int_{|x| \leq r} \rho (x) \, dx = 4\pi \int_0^r s^2 \rho (s) \, ds.  \label{defmrho} \\
U_{\rho}'(r) &:= \frac{m_{\rho}(r)}{r^2} \\
\nabla U_{\rho} (x) &:= \frac{m_{\rho} (r)}{r^2} \, \frac{x}{r}  \label{nablau}\\
U_{\rho} (r) &:= - \int _r^\infty U_{\rho}'(s) \, ds
\end{align}
and we will sometimes write $U_f$ or $\nabla U_f$ instead of $U_{\rho_f} = U_\rho$ or $\nabla U_{\rho_f} = \nabla U_{\rho}$, if $\rho _f $ is induced by $f$.
The definition (\ref{nablau}) implies
$$ \int_{\mathbb{R}^3} |\nabla U_{\rho _f}(x)|^2 \, dx = 4\pi\int_0^{\infty} \frac{m^2_{\rho} (r)}{r^2} \, dr.$$
Now we can state the next lemma.
\begin{lemma} \label{rhoestimates}
\begin{itemize}
\item[(a)] Define the function $\zeta \in C(\mathbb{R}^+)$ by
$$ \zeta(R) = \left\{ \begin{array}{ccl} R^{q_1} & \text{for} & 0\leq R \leq 1 \\
R^{q_2} &\text{for} & 1<R<\infty  \end{array} \right.,$$
where $q_1:= l-k+1/2>0$ and $q_2:= 4l+5-n>0$. Then there exists a constant $C>0$ such that for $\rho \in L^{n,l} (\mathbb{R}^3)$ with 
$ \int \rho (x) dx = M$ we have
\begin{align*}
-\epot (\rho)&:= \frac{1}{8\pi} \int |\nabla U_{\rho}|^2 dx + \int \frac{M_c}{|x|}\, \rho (x) dx\\ 
&\leq \frac{1}{2} \int_0^R \left( \frac{m^2_{\rho} (r)}{r^2} + 8\pi M_c \, r \rho(r) \right)dr + \frac{1}{2R}\left(M^2+2MM_c\right) \\
&\leq C\zeta (R)(1+ \| \rho \|_{n,l}^{1+\frac{1}{n}}) + \frac{1}{2R}\left( M^2+2MM_c \right), \qquad R>0
\end{align*}
where $U_{\rho} $ denotes the potential induced by $\rho$.
\item[(b)] For every $R>0$ the mapping
$$ T: L^{n,l}(\mathbb{R}^3) \ni \rho \mapsto \frac{m_{\rho}}{r} \lvert_{[0,R]} \, \,  \in L^2([0,R]) $$
is compact.
\item[(c)] For $\rho_1, \rho_2 \in L^{n,l} (\mathbb{R}^3) \cap L^1 (\mathbb{R}^3)$ we have
\[
\int \nabla U_{\rho_1} \cdot \nabla U_{\rho_2} \, dx = - 4\pi \int  U_{\rho_1} \rho_2 \, dx.
\]
\end{itemize}
\end{lemma}
\prf
Obviously, we have $m_{\rho} (r) \leq M$, and this shows the first estimate of (a). Now for $\rho \in L^{n,k}(\mathbb{R}^3)$, we have
\begin{align*}
\int_0^R r\rho(r) dr  &= \int_0^R  r^{\frac{l-k-1/2}{n+1}} r^{\frac{2k+3}{n+1}} \rho(r) \, dr  \\
&\leq \bigg( \int_0^R r^{l-k-1/2} \, dr \bigg) ^{\frac{1}{n+1}} \bigg( \int_0^R r^{\frac{2k+3}{n}} \rho^{1+1/n}(r) \, dr \bigg) ^{\frac{n}{n+1}} \\
&\leq CR^{l-k+1/2} \|\rho\|_{n,l} \\
&\leq CR^{l-k+1/2}(1+ \| \rho \|_{n,l}^{1+\frac{1}{n}})
\end{align*}
where we used H\"older's inequality in the second line. Furthermore, again by H\"older's inequality,
\begin{equation} \label{mrhoest}
|m_{\rho} (r)| \leq C r^{(2l+3)/(n+1)}\|\rho \|_{n,l}, \quad r\geq 0, 
\end{equation}
and thus
\begin{align}
\int_0^R \frac{m^2_{\rho} (r)}{r^2} dr &\leq C\|\rho \|_{n,l}^2 R^{(4l+5-n)/(n+1)} \label{bounded} \\
&\leq C  R^{(4l+5-n)/(n+1)}(1+ \| \rho \|_{n,l}^{1+\frac{1}{n}}), \notag
\end{align}
which implies the estimate in (a). As to (b), by (\ref{bounded}) we already know that the operator $T$ is bounded. To show the compactness of $T$, we use the Fr\'echet-Kolmogorov criterion, cf. \cite{Yosida}, Theorem X.1. We take a bounded set $K \in L^{n,l}$ and to show the precompactness of $TK$, we redefine $T\rho:= \frac{m_{\rho}}{r}\chi _{[0,R]} \in L^2(\mathbb{R})$. The crucial part is to show that
$$ \| (T\rho)_h - T\rho \|_2 \rightarrow 0, \quad h\rightarrow 0, $$
uniformly in $\rho \in K$, where $(T\rho)_h := (T\rho)(\cdot+h)$.
For $h>0$, we have
\begin{align*}
 \bigg\| \frac{m_{\rho} (r+h)}{r+h} &\chi _{[0,R]}(r+h) - \frac{m_{\rho} (r)}{r}  \chi _{[0,R]}(r)\bigg\|_2^2 \\
&\leq \int_0^h \frac{m^2_{\rho} (r+h)}{r^2}\, dr + \int_0^h \frac{m^2_{\rho} (r)}{r^2}\, dr + \int_{R-h}^R \frac{m^2_{\rho} (r)}{r^2} \, dr \\
& \hspace{8mm} + \int_h^{R-h} m_{\rho}^2(r) \bigg| \frac{1}{r+h} - \frac{1}{r} \bigg|^2 \, dr \\
& \hspace{8mm} + \int_h^{R-h} \frac{1}{(r+h)^2} \big| m_{\rho}(r+h) - m_{\rho}(r) \big|^2 \, dr 
\end{align*}
For the first four terms, one can use the estimate (\ref{mrhoest}). Indeed, for example,
\begin{align*}
\int_{R-h}^R \frac{m^2_{\rho} (r)}{r^2} \, dr &\leq C \int_{R-h}^R \| \rho \|_{n,l}^2 r^{(4l+4-2n)/(n+1)} \, dr \\
&= C\| \rho \|_{n,l}^2 \left( R^{(4l+5-n)/(n+1)}- (R-h)^{(4l+5-n)/(n+1)} \right) \\
\end{align*}
and
$$ \int_h^{R-h} m_{\rho}^2(r) \bigg| \frac{1}{r+h} - \frac{1}{r} \bigg|^2 \, dr = \int_h^{R-h} \frac{m_{\rho}^2(r)}{r^2} \left( \frac{h}{r+h} \right)^2 \, dr, $$
which converges to zero by Lebesgue's theorem. We have
$$ | m_{\rho}(r+h) - m_{\rho}(r)| \leq C \|\rho \|_{n,l} \left( (r+h)^{(2l+3)/(n+1)} - r^{(2l+3)/(n+1)} \right), $$
and again by Lebesgue's theorem, also the last term converges to zero. Each term coneverges uniformly in $\rho \in K$ and the case $h<0$ is completely analoguous. \\
As to (c), we firstly show the assertion for $\rho_1, \rho_2 \in C^{\infty}\cap L^{n,l}\cap L^1$. An integration by parts gives
\begin{align*}
\int \nabla U_{\rho_1} \cdot \nabla U_{\rho_2} \, dx &= 4\pi \int_{\mathbb{R}^+} U_{\rho _1}'(r) m_{\rho_2} (r) \, dr \\
&= 4\pi U_{\rho_1}(r) m_{\rho_2} (r) \bigg|_{r=0}^{r=\infty} - (4\pi)^2 \int_{\mathbb{R}^+} U_{\rho _1} (r) r^2 \rho_2 (r) \, dr \\
&= - 4\pi\int  U_{\rho _1} \rho _2 \, dx,
\end{align*}
where the boundary term at infinity vanishes since $|U_{\rho _1}  (r)| \leq \| \rho _1\|_1 / r$ and $m_{\rho _2} (r) \leq \|\rho _2\|_1$ and the boundary term at zero vanishes since $m_{\rho _2} (r) = O(r^2), \enspace r\rightarrow 0$.
Now we consider approximating sequences $(\rho_1^j),(\rho_2^j) \subset L^{n,l} \cap C^{\infty} \cap L^1$ such that for $i=1,2$
$$ \rho_i^j \rightarrow \rho _i \quad \text{in} \enspace L^{n,l} \quad (j \rightarrow \infty), $$
and $\| \rho_i^j \|_1 \leq \|\rho_i\|_1$. Using the estimates of (a), we conclude that the above identity still holds for $\rho _i \in L^{n,l}\cap L^1$ and the proof is complete.
\prfe
\begin{lemma} \label{lowerbound}
There exists a constant $C>0$, such that
$$ \Hc (f) \geq \frac{1}{2} \left( \ekin (f) + \mathcal{C} (f) \right) - C, \quad f \in \mathcal{F}_M$$
in particular,
\be
\label{defhm}
h_M:= \inf \{ \Hc (f) \, | \, f\in \mathcal{F} _M \} > -\infty.
\ee
\end{lemma}
\prf
Using the previous two lemmas we have
\begin{align*}
\Hc (f) &\geq \ekin (f) + \mathcal{C} (f) - C\zeta(R)(1+ \| \rho _f \|_{n,l}^{1+\frac{1}{n}} ) - \frac{M^2+2MM_c}{2R} \\
&\geq (\ekin(f) + \mathcal{C} (f))(1-C\zeta(R)) - C\zeta(R) - \frac{M^2+2MM_c}{2R},
\end{align*}
where $C>0$ is some constant which does not depend on $R>0$. The assertion follows by a suitable choice of $R$.
\prfe
\section{A scaling lemma}
\setcounter{equation}{0}
In this section we show that $h_M$ is negative. We also examine the behaviour of $\Hc (f)$, if $f$ is rescaled.
\begin{lemma} \label{scaling} 
Define $h_M$ as in (\ref{defhm}).
Then for $M>0$ we have $-\infty < h_M <0$.
\end{lemma}
\prf
As already mentioned in the introduction, we will use coordinates adapted to spherical symmetry. If $f(x,v) = f(Ax,Av) \enspace \forall \enspace A \in O(3)$, we have
\[ f(x,v) = \underline{f} (r,w,L),\] 
where $r:= |x|, \enspace w:= \frac{x\cdot v}{r}, \enspace L:= | x\times v |^2 $ and we will write again $f$ instead of $\underline{f}$. 

It is easy to check that, in the new coordinates, the energies and the Casimir functional read
\begin{align*}
 \ekin (f) &= 2\pi^2 \int_{\mathbb{R}^+} \int_{\mathbb{R}} \int_{\mathbb{R}^+} ( w^2 + \frac{L}{r^2} ) f(r,w,L) \, dLdwdr, \\
\epot (f) &= -\frac{1}{2} \int_{\mathbb{R}^+} \frac{m_f^2(r)}{r^2} dr- 4\pi M_c\int_{\mathbb{R}^+} r \rho_f(r) dr, \\
\mathcal{C}(f) &=  {4\pi^2} \int_{\mathbb{R}^+} \int_{\mathbb{R}} \int_{\mathbb{R}^+} f^{1+1/k}(r,w,L) (L-L_0)_+^{-l/k} \, dLdwdr,
\end{align*}
with $\mathbb{R}^+ := [0,\infty[$ and $m_f = m_{\rho_f}$ as in (\ref{defmrho}).

Given any function $f \in \mathcal{F}_M$, we define a rescaled and translated function 
\begin{equation} \label{defbarf}
\bar{f}(r,w,L) = af\left(br,cw,b^2c^2L-(b^2c^2-1)L_0 \right),
\end{equation}
where $a,b,c >0$. \\
Then $\bar{f} (r,w,L) = 0$ a.e. if $L < L_0$,
\[ 
\iiint \bar{f} (r,w,L)\, dLdwdr = a(bc)^{-3} \iiint f(r,w,L) \,dL dw dr 
\]
and if $f\in \mathcal{F} _M$, we have $\bar{f} \in \mathcal{F}_{\bar{M}}$ with $\bar{M}= a(bc)^{-3}M$.
Furthermore,
\begin{align}
 \ekin(\bar{f}) &= 2\pi^2 a b^{-3} c^{-5} \iiint \left(w^2 + \frac{L + (b^2c^2-1)L_0}{r^2} \right) f\left(r,w,L \right)\, dr dw dL, \label{ekinbarf} \\
\mathcal{C} (\bar{f}) &= a^{1+\frac{1}{k}} b^{-3+\frac{2l}{k}} c^{-3+\frac{2l}{k}} \mathcal{C} (f),\label{cbarf} \\
\epot (\bar{f}) &= -\frac{1}{2} \int_{\mathbb{R}^+} a^2b^{-6}c^{-6}\frac{m_f^2(br)}{r^2} dr - 4\pi ab^{-2}c^{-3}\int_{\mathbb{R}^+} M_cr\rho_f(r)dr \notag \\
&= -\frac{1}{2} a^2b^{-5}c^{-6} \int_{\mathbb{R}^+} \frac{m_f^2(r)}{r^2} dr - 4\pi M_c ab^{-2}c^{-3}\int_{\mathbb{R}^+} r\rho_f(r)dr. \label{epotbarf}
\end{align}
To prove the lemma, we consider the case $bc < 1$. Here we have
 \be \label{ekinab}
 \ekin(\bar{f}) \leq  ab^{-3}c^{-5} \ekin (f).
 \ee
 Now we fix some $f \in \mathcal{F}_1$ with compact support and let
 $$ a = M (bc)^3. $$
Consequently,
 \begin{align*}
 \Hc (\bar{f}) &\leq a^{1+\frac{1}{k}} b^{-3+\frac{2l}{k}} c^{-3+\frac{2l}{k}} \mathcal{C} (f) + ab^{-3}c^{-5} \ekin (f) \\
 &\hspace{5mm}-\frac{1}{2} a^2b^{-5}c^{-6} \int_{\mathbb{R}^+} \frac{m_f^2(r)}{r^2} dr - 4\pi M_c ab^{-2}c^{-3}\int_{\mathbb{R}^+} r\rho_f(r)dr \\
 &\leq C_1 a^{\frac{1}{k}} (bc)^{\frac{2l}{k}} + C_2 c^{-2} - C_3 b,
 \end{align*}
 where $C_1,C_2,C_3 >0$ depend on $f$ and $M$. Since we want the last term to dominate as $b \rightarrow 0$, we let
 $c= b^{-\eta /2}$, so that $bc=b^{1- \frac{\eta}{2}}$ for some $\eta \in ]1,2[$. For $b$ small enough we have $bc < 1$ and
$$\Hc (\bar{f}) \leq C_1 b^{(1-\frac{\eta}{2})(2l+3)/k} + C_2 b^{\eta} - C_3 b. $$
Now fix $\eta \in ]1,2[$ such that $(1-\frac{\eta}{2})(2l+3)/k > 1$; such an $\eta$ exists by the assumptions on $k$ and $l$. For $b>0$ sufficiently small, the sum of the last three terms will be negative and the assertion follows.
\prfe
In the next section, we will use the rescaling formulas (\ref{ekinbarf})--(\ref{epotbarf}) to show that a function $f_0$, constructed by the weak limit of a minimizing sequence actually is a minimizer with mass $M$. 
\section{Existence and properties of minimizers}
\setcounter{equation}{0}
\begin{theorem} \label{minihc}
Let $M>0$, $L_0 >0$ and let $(f_j) \subset \mathcal{F}_M$ be a minimizing sequence of $\Hc$. Then there is a minimizer $f_0$ and a subsequence $(f_{j_k})$ such that $\Hc (f_0) = h_M$ and
$f_{j_k} \rightharpoonup f_0$ weakly in $L^{k,l}$. For the induced potentials we have $\nabla U_{j_k} \rightarrow \nabla U_0$ strongly in $L^2(\mathbb{R}^3)$.
\end{theorem}
\prf
 By Lemma \ref{lowerbound}, $\ekin (f_j) + \mathcal{C} (f_j) $ is bounded and thus $(f_j)$ is bounded in $L^{k,l}$. Now there exists a weakly convergent subsequence, denoted by $(f_j)$ again:
\[
f_j \rightharpoonup f_0 \quad \text{weakly in} \enspace L^{k,l}.
\]
 Clearly, $f_0 \geq 0$ a.e. and $f_0 (x,v) = 0$ a.e. for $0\leq L < L_0$.
 By weak convergence, 
 \begin{equation} \label{ekin}
 \ekin (f_0) \leq \limsup_{j\rightarrow \infty} \ekin (f_j) < \infty.
 \end{equation}
 By Lemma \ref{lemmarhobound}, $(\rho _j) = (\rho _{f_j})$ is bounded in $L^{n,l} (\mathbb{R}^3)$. After choosing another subsequence, we conclude that
\begin{equation} \label{weakrho}
 \rho _j \rightharpoonup \rho _0 \quad \text{weakly in} \enspace L^{n,l},
\end{equation}
where we have the identity
$$ \rho_0 = \rho _{f_0} := \int f_0 (x,v) \, dv. $$
Indeed, assume we would have $\rho _{f_0} > \rho _0$ a.e. on the measurable set $A:=A_{R_1,R_2} := \{ x \in \mathbb{R}^3 \, | \, R_1 < |x| < R_2$ with $0<R_1<R_2 < \infty \}$, note that both $\rho _0$ and $\rho _{f_0}$ are spherically symmetric. Then for $R>0$, by weak convergence we have
\begin{align*}
0< \gamma &:= \int_A (\rho_{f_0}(x) - \rho _0 (x) )\, dx \\
&=  \lim _{j\rightarrow \infty} \int_A \int _{|v|<R}  f _j (x,v) \, dvdx  + \int_A \int _{|v|>R}  f _0 (x,v) \, dvdx -\\
& \hspace{4mm} - \lim _{j\rightarrow \infty} \int_A \rho _j (x) \, dx,
\end{align*}
where we used the fact that $\chi _A \in \big(L^{n,l}\big) ^*$ and $\chi _{A\times B_R} \in \big( L^{k,l} \big) ^*$. Now $\ekin (f_j)$ is bounded and this implies
$$ \int_A  \int _{|v|>R}  f _0 (x,v) \, dvdx \leq \frac{2}{R^2} \ekin (f_0) \leq  \frac{2}{R^2}\limsup _{j\rightarrow \infty} \ekin (f_j) \leq \frac{C}{R^2}. $$
We conclude
$$ |\gamma| \leq \frac{C}{R^2} + \lim _{j\rightarrow \infty}  \int_A \int _{|v|>R}  f _j (x,v) \, dvdx \leq \frac{2C}{R^2}, $$
which is a contradiction.

Next, from (\ref{weakrho}) together with Lemma \ref{rhoestimates} (a) (b), the strong convergence 
 \begin{equation} \label{epot}
 \nabla U_j \rightarrow \nabla U _0 \quad \text{strongly in} \enspace L^2(\mathbb{R}^3),
  \end{equation}
follows,
and we have 
\[
\epot (f_j) \rightarrow \epot (f_0).
\]
Indeed, from Lemma \ref{rhoestimates} we have
\begin{align*}
\frac{1}{4\pi} \int | \nabla U_j - \nabla U_0 |^2 \, dx &= \int_0^{\infty} \frac{m^2_{\rho _{f_j} - \rho _{f_0}}}{r^2} \, dr \\
&\leq \int_0^R \frac{m^2_{\rho_{f_j} - \rho_{f_0}}}{r^2} \, dr + \frac{M^2}{R} =: I + II .
\end{align*}
Now let $\epsilon >0$ be given. Choose $R>0$ large enough so that $II < \epsilon /2$. For $j$ sufficiently large, the first term also will be smaller than $\epsilon /2$ because of the compactness of $T$, defined in Lemma \ref{rhoestimates} (b): The weak convergence $\rho _{f_j} \rightharpoonup \rho _{f_0}$ implies the strong convergence $m_{\rho_{f_j} - \rho_{f_0}}/r  \rightarrow 0$ in $L^2([0,R])$.

Furthermore, we can estimate the interaction term as
$$ \bigg| \int_{\mathbb{R}^3}  \frac{1}{|x|} \left( \rho _j (x) - \rho _0(x) \right)  \, dx \bigg|  \leq  \bigg|  \int_{B_R}  \frac{1}{|x|} \left( \rho _j (x) - \rho _0(x) \right)  \, dx \bigg| + \frac{2M}{R}. $$
Here the first term tends to zero, because of the weak convergence (\ref{weakrho}) together with the fact that $\langle \frac{1}{|x|},\cdot \rangle_{L^2(B_R)} \in \left( L^{n,l} (\mathbb{R}^3) \right)^*$ which we have shown in the proof of Lemma \ref{rhoestimates}(a). The same
argument as above then proves
$$ \bigg| \int_{\mathbb{R}^3}  \frac{1}{|x|} \left( \rho _j (x) - \rho _0(x) \right)  \, dx \bigg|  \rightarrow 0.$$

Next, we show that $f_0$ actually is a minimizer, in particular \\
$\ekin (f_0) + \mathcal{C} (f_0) < \infty$. By weak covergence, we have
 $$ \mathcal{C} (f_0) = \| f_0 \|_{k,l}^{(k+1)/k} \leq \liminf_{j\rightarrow \infty} \| f_j \| _{k,l}^{(k+1)/k} < \infty. $$
 Together with (\ref{ekin}) and (\ref{epot}) this implies
 $$ \ekin(f_0) + \mathcal{C} (f_0) \leq \lim_{j\rightarrow \infty} (\ekin(f_j) + \mathcal{C} (f_j)) < \infty, $$
 note that the $\lim_{j\rightarrow \infty}$ in the above inequality exists. Finally,
 $$ \Hc (f_0) = \mathcal{C} (f_0) + \ekin (f_0) + \epot (f_0) \leq \lim_{j\rightarrow \infty} \left( 
  \mathcal{C} (f_j) + \ekin (f_j) + \epot (f_j) \right) = h_M. $$
It remains to show that $\|f_0\|_1 = M$. By weak convergence, we have $\|f_0\|_1 \leq M$ and we already know that $\|f_0\|_1>0$, since $h_M<0$. Now assume that $M_0:= \|f_0\|_1 <M$. We consider the rescaled function $\bar{f_0}$ defined in (\ref{defbarf}) in section 4 and recall formulas (\ref{ekinbarf})--(\ref{epotbarf}). Now define 
$$ a:=1, \quad c:= \left( \frac{M_0}{M} \right)^{-1/3}, \quad b:= c^{-2}. $$
This implies $(bc)^{-3} = M/M_0$ and thus $\|\bar{f_0}\|_1 = M$. We have
\begin{align*}
h_M &\leq \Hc(\bar{f_0}) \\
& \leq c \ekin (f_0) + c^{3-2l/k} \mathcal{C} (f_0) - \frac{1}{2} c^4 \int_{\mathbb{R}^+} \frac{m_f^2(r)}{r^2} dr - c 4\pi M_c \int_0^{\infty} r\rho_f(r)dr,
\end{align*}
where we used (\ref{ekinab}), note that $bc=c^{-1} <1$. Since $c>1$ and $0<k\leq l$ we conclude
\be \label{scarg}
h_M \leq \Hc(\bar{f_0}) \leq  c \Hc (f_0) = \left( \frac{M}{M_0} \right) ^{1/3} h_M,
\ee
which is a contradiction.
\prfe
\begin{theorem} \label{statel}
Let $f_0 \in \mathcal{F}_M$ be a minimizer of $\Hc$. Then there exists $E_0 <0$ such that
\be \label{f0form}
f_0 (x,v) =  \frac{k}{k+1}(E_0 - E)_+^k(L-L_0)_+^l
\ee
where
\be \label{micenergy}
E:= \frac{1}{2} v^2 + U_0(x)- \frac{M_c}{|x|}
\ee
and $U_0$ is the potential induced by $f_0$. Moreover, $f_0$ is a steady state of the Vlasov-Poisson system (\ref{poisson})--(\ref{vlasov}).
\end{theorem}
\prf
Let $f_0$ be a minimizer. We choose a suitable representative for $f_0$ and define
for $\epsilon >0$ the set
$$ K_{\epsilon} := \left\{ (x,v) | \epsilon < f_0(x,v) \leq \frac{1}{\epsilon}, \enspace L_0 + \epsilon \leq L \leq L_0 + \frac{1}{\epsilon} \right\}. $$
Since $f_0 \in L^{k,l}$ we have $0<|K_{\epsilon}|< \infty$ for $\epsilon$ sufficiently small. 
Now let $g \in L^{\infty} (\mathbb{R}^6)$ be spherically symmetric with supp $g \subset K_{\epsilon}$, and
\[
h:= g - \frac{1}{|K_{\epsilon}|} \left( \iint g \, dvdx \right) \cdot \chi _{K_{\epsilon}}.
\]
Then for $\tau \in \mathbb{R}$ small enough we have $f_0 + \tau h \geq 0$ and $f_0 + \tau h \in \mathcal{F}_M$, indeed, $\ekin (f_0 + \tau h) < \infty$ and
$$ \mathcal{C} (f_0 + \tau h) = \mathcal{C} (f_0) + \tau \, \iint \Phi'(f_0) (L-L_0)_+^{-\frac{l}{k}}h + o(\tau) < \infty,$$
where we recall that $\Phi (f) = f^{1+1/k}$. Now we have
\begin{align*}
0 &\leq \Hc (f_0 + \tau h) - \Hc (f_0) =\\
&\hspace{16mm}= \tau \, \iint \left( \Phi'(f_0) (L-L_0)_+^{-l/k} + \frac{1}{2} v^2 + U_0 (x) -\frac{M_c}{|x|} \right) h \, dvdx + o(\tau) \\
&\hspace{16mm}= \tau \,\iint \left( \Phi'(f_0) (L-L_0)_+^{-l/k} + E \right) h \, dvdx  + o(\tau),
\end{align*}
where we used Lemma \ref{rhoestimates} (c) to calculate the potential energy term.
Since $-h$ is also an admissible function, this implies
$$ \iint \left( \Phi'(f_0) (L-L_0)_+^{-l/k} + E \right) h \, dvdx = 0.$$
Inserting the definition of $h$ we get
$$ \iint \left[ \left( \Phi'((L-L_0)_+^{-l}f_0) + E \right) - \frac{1}{|K_{\epsilon}|} \iint_{K_{\epsilon}} \left( \Phi'((L-L_0)_+^{-l}f_0) + E \right) \right] g  \, dvdx = 0. $$
Consequently,
$$ \Phi'((L-L_0)_+^{-l}f_0) + E = E_{\epsilon} \quad \text{a.e. on} \enspace K_{\epsilon}, $$
where
$$ E_{\epsilon} := \frac{1}{|K_{\epsilon}|} \iint_{K_{\epsilon}} \left( \Phi'((L-L_0)_+^{-l}f_0) + E\right) \, dvdx. $$
Thus for $\epsilon$ small, $E_{\epsilon}$ will be a constant which we denote by $E_0$ and we conclude
\begin{equation}  \label{eulerlagr}
\Phi'((L-L_0)_+^{-l}f_0) + E = E_0 \quad \text{a.e. on} \enspace \{(x,v)| f_0 (x,v) > 0\}.
\end{equation}
Suppose now, there would exist a measurable set $A \subset \{(x,v)| f_0 (x,v) = 0, L_0 \leq L\}$ with 
$$ E < E_0 \quad \text{a.e. on} \enspace A $$
and $0 < |A| < \infty$. We can also assume that $A$ is spherically symmetric, i.e. $\chi _A$ is spherically symmteric. Next, define
$$ h:= \chi_A - \frac{1}{|K_{\epsilon}|} \left( \iint \chi_A dvdx \right) \cdot \chi _{K_{\epsilon}} $$
with $K_{\epsilon}$ as above and small $\epsilon > 0$. Then for $\tau >0$ sufficiently small we have
$f_0 + \tau h \in \mathcal{F}_M$ and again
\begin{equation*}
0 \leq \Hc (f_0 + \tau h) - \Hc (f_0) 
= \tau \,\iint \left( \Phi'((L-L_0)_+^{-l}f_0) + E \right) h \, dvdx  + o(\tau).
\end{equation*}
Plugging the definition of $h$ into the above equation, we have
\begin{align*}
0 &\leq \iint  \left( \Phi'((L-L_0)_+^{-l}f_0) + E \right) \chi_A - E_0 \iint  \chi_A \\
  &= \iint_A (E-E_0) < 0,
\end{align*}
a contradiction and thus $E \geq E_0$ a.e. on $\{(x,v)| f_0 (x,v) = 0, L_0 \leq L\} $. Together with (\ref{eulerlagr}) this implies that $f_0$ is of the form given in the theorem. 

Since $f_0$ is a function of the microscopic energy $E$ defined by (\ref{micenergy}) and $L$, it is constant along solutions of the characteristic system
\begin{equation*}
\left\{ \begin{array}{rcl} \dot{X} &= &V \\ \dot{V} &= & - \nabla _x U_0(X)- \frac{M_c}{|X|^3} X  \end{array} \right.
\end{equation*}
and thus $f_0$ is a solution of the Vlasov equation, provided the potential $U_0$ is sufficiently smooth. But one can indeed show that $U_0 \in C^2(\mathbb{R}^3)$. This can be seen as follows. We firstly recall the formula for $\rho_{f_0}$, if $f_0$ is of form (\ref{f0form}),
\be \label{rho0equ}
\rho_0 (r) :=  \rho _{f_0}(r) = C(k,l) \, r^{2l} \, \bigg( E_0 - U_0(r) + \frac{M_c}{r} - \frac{L_0}{r^2} \bigg)_+^{k+l+3/2} 
\ee
and we claim that $U_0  \in L^{\infty}(\mathbb{R^+})$ and thus the above equation implies $\rho_{f_0} \in L^1 \cap L^{\infty}$. Indeed, for any $R>r$,
\begin{align*}
-U_0(r) &= \int_r^R \frac{m_{\rho _0}(s)}{s^2} \, ds + \int_R^{\infty} \frac{m_{\rho_0}(s)}{s^2} \, ds \\
&\leq C\int_r^R s^{(-2k-2)/(n+1)} \|\rho_0\|_{n,l} \, ds + \frac{M}{R} \\
&= C \|\rho_0\|_{n,l} \big( R^{(-k+l+1/2)/(n+1)} - r^{(-k+l+1/2)/(n+1)} \big) + \frac{M}{R},
\end{align*}
and because of $0<k<l+1/2$, the claim follows. Now $\rho_{f_0} \in L^1 \cap L^{\infty}$ implies $U_0 \in C^1$ and because of (\ref{rho0equ}) also $\rho _0 \in C^1$. Together with $U_0'(r) = \frac{1}{r^2} \int _0^r s^2\rho_0(s) \, ds $, the asserted regularity of $U_0$ is proved. 

By construction, we have
$$ \Delta U_0= 4\pi \rho _0, $$
so that $(f_0, \rho_0 , U_0)$ is indeed a solution of the Vlasov-Poisson system. It remains to show that $E_0<0$. Recall the formula for $\rho _0$ from (\ref{rho0equ}) and the fact that $\|f_0 \|_1 = M$. If $E_0 \geq 0$, we would have
$$ \|f_0\| _1 = \|\rho _0\|_1 \geq C(k,l)  \int_{R_0}^{\infty} r^{2l+2} \bigg( \frac{M_c}{2r} \bigg) ^{k+l+3/2}\,dr = C \int_{R_0}^{\infty} r^{l-k+1/2} \, dr = \infty,$$
where we have chosen $R_0 >0$ sufficiently large so that $L_0/r^2 < M_c/2r, \enspace r>R_0$. Consequently, we conclude $E_0 < 0$.
\prfe
\section{Dynamical stability}
\setcounter{equation}{0}
We investigate the nonlinear stability of $f_0$. For $f \in \mathcal{F}_M$,
\begin{equation}
\Hc(f) - \Hc(f_0) =
 d(f,f_0) - \frac{1}{8\pi} \int |\nabla U_f-\nabla U_{f_0}|^2 dx \label{diffest},
\end{equation}
where 
\begin{equation*}
d(f,f_0) := \iint \left[ \left( f^{1+1/k} - f_0^{1+1/k} \right) (L-L_0)_+^{-l/k}+ (E-E_0) (f-f_0) \right] \, dvdx,
\end{equation*}
where $E$ is defined as in (\ref{micenergy}). We have $d(f,f_0) \geq 0, \enspace f \in \mathcal{F}_M$ with $d(f,f_0) = 0$, iff $f=f_0$. Indeed,
\[
d(f,f_0) \geq \iint \left[  \Phi'((L-L_0)_+^{-l}f_0)  + (E-E_0) \right] (f-f_0)\, dvdx \geq 0,
\]
which is due to the convexity of $\Phi$, and on the support of $f_0$ the bracket vanishes. This fact allows us to use $d(.,f_0)$ to measure the distance to the stationary solution $f_0$.
\begin{theorem}  \label{stability}
Assume that the minimizer $f_0$ is unique in $\mathcal{F}_M$. Then for all $\epsilon >0$ there is $\delta >0$ such that for any solution $f(t)$ of the Vlasov-Poisson system with $f(0) \in 
C^1_c (\mathbb{R}^6) \cap \mathcal{F}_M$,
$$ d(f(0),f_0) + \frac{1}{8\pi} \int |\nabla U_{f(0)}-\nabla U_{f_0}|^2 dx < \delta $$
implies
$$ d(f(t),f_0) + \frac{1}{8\pi} \int |\nabla U_{f(t)}-\nabla U_{f_0}|^2 dx < \epsilon, \quad t\geq 0. $$
\end{theorem}
\prf
We observe that $\Hc$ is conserved along any solution $f(t)$ of the Vlasov-Poisson system with  $f(0) \in 
C^1_c (\mathbb{R}^6) \cap \mathcal{F}_M$. This follows from conservation of energy and the fact that both $f(t)$ and $L$ are conserved along characteristics. Assume the theorem were false. Then there exists $\epsilon _0 >0, \quad t_j >0$, and $f_j(0) \in 
C^1_c (\mathbb{R}^6) \cap \mathcal{F}_M$ such that
$$ d(f_j(0),f_0) + \frac{1}{8\pi} \int |\nabla U_{f_j(0)}-\nabla U_{f_0}|^2 dx \leq \frac{1}{j} $$
and
$$ d(f_j(t_j),f_0) + \frac{1}{8\pi} \int |\nabla U_{f_j(t_j)}-\nabla U_{f_0}|^2 dx \geq \epsilon_0. $$
From (\ref{diffest}), we have
$$ \lim_{j\rightarrow \infty} \Hc (f_j(0)) = h_M ,$$ and because $\Hc (f_j (t))$ is conserved,
$$ \lim_{j\rightarrow \infty} \Hc (f_j(t_j)) = \lim_{j\rightarrow \infty} \Hc (f_j(0)) = h_M. $$
Thus $(f_j(t_j))  \subset \mathcal{F}_M $ is a minimizing sequence of $\Hc$ and with Theorem \ref{minihc} we have
$$ \frac{1}{8\pi} \int |\nabla U_{f_j(t_j)}-\nabla U_{f_0}|^2 dx\rightarrow 0,$$
which implies
$$ d(f_j(t_j),f_0) \rightarrow 0 $$
by (\ref{diffest}), a contradiction.
\prfe
\begin{corollary}
If in Theorem \ref{stability} the assumption $\|f(0)\|_{k,l} = \|f_0\|_{k,l}$ is added, then for any $\epsilon >0$ the parameter $\delta >0$ can be chosen such that the stability estimate
\[
\| f(t) - f_0 \|_{k,l} < \epsilon, \quad t\geq 0
\]
holds.
\end{corollary}
\prf
We repeat the proof of Theorem \ref{stability} except that in the contradiction assumption have
\[
\| f_j(t_j) - f_0 \|_{k,l} +d(f_j(t_j),f_0) + \frac{1}{8\pi} \int |\nabla U_{f_j(t_j)}-\nabla U_{f_0}|^2 dx \geq \epsilon_0. 
\]
From the minimizing sequence $f_j(t_j)$ we can now extract a subsequence which converges weakly in $L^{k,l}$ to $f_0$. But due to our additional restriction we have $$\|f_j(t_j)\|_{k,l} = \|f_0\|_{k,l}, \quad j \in \mathbb{N}.$$ 
Now the lower semicontinuity of the norm and the uniform convexity of $L^{k,l}(\mathbb{R}^6)$ imply
$f_j (t_j) \rightarrow f_0 $ strongly in $L^{k,l}$. Together with the rest of the proof of Theorem \ref{stability}, the assertion follows.
\prfe
\newpage
\noindent
\textbf{Remarks.}
\begin{itemize}
\item[(a)] The technical assumption $f= 0$ a.e. for $0 < L < L_0$ in the class of perturbations $\mathcal{F}_M$, see (\ref{defoffm}), is needed for the scaling argument in Lemma \ref{scaling} and it would be desirable to improve it to $f= 0$ a.e. for $0 < L < \gamma L_0$ for some $0<\gamma <1$.
\item[(b)] For $M_c=0$ one can show existence and stability for steady states of form (\ref{ansatz}) for the parameter range $l>-1, \enspace 0<k<l+3/2$, see \cite{SCH}. For $M_c>0$, we had to restrict the parameter range to $0<k\leq l$ in order to guarantee that the scaling argument (\ref{scarg}) works.
\item[(c)] The uniqueness of the minimizer $f_0$ subject to the fixed mass constraint can be shown by a scaling argument in the case
$L_0 = 0$ and $M_c =0$. For $L_0 >0$, at least numerically the minimizer seems to be unique, but the scaling argument fails because of the translation in $L$. We mention that, for Theorem \ref{stability}, it would suffice if the minimizers of $\Hc$ were isolated.
\item[(d)] We only obtain stability against spherically symmetric perturbations, because the quantity $L$
is conserved by the characteristic flow only for spherically symmetric solutions. Stability against asymmetric perturbations is an open problem and more delicate mathematical tools have to be invented to address this question.
\end{itemize}
\textbf{Acknowledgements.}
The author wishes to thank Gerhard Rein for the critical review of the manuscript. This research was supported by the Deutsche Forschungsgemeinschaft under the project ``Nichtlineare Stabilit\"at bei ki\-ne\-ti\-schen Modellen aus der Astrophysik und Plasmaphysik''.

\bibliography{shells_bh}{}
\bibliographystyle{plain}
\end{document}